\documentclass[pra,twocolumn,letterpaper,showkeys,showpacs, superscriptaddress]{revtex4-1}
\usepackage{graphicx,amsmath,amssymb,amsfonts,latexsym,color,dcolumn,bm}
\usepackage{gensymb}
\usepackage{braket}

\begin{document}

\newcommand{\be}{\begin{equation}}
\newcommand{\ee}{\end{equation}}
\newcommand{\bea}{\begin{eqnarray}}
\newcommand{\eea}{\end{eqnarray}}
\newcommand{\ra}{\rangle}
\newcommand{\la}{\langle}
\newcommand{\om}{\omega}

\title{Effect of quantum and thermal jitter on the feasibility of Bekenstein's proposed experiment to search for Planck-scale signals}
\author{G. Jordan Maclay}
\affiliation{Quantum Fields LLC, 147 Hunt Club Dr., St. Charles, IL. 60174}
\email{jordanmaclay@quantumfields.com}
\author{S. A. Wadood}
\affiliation{Institute of Optics, University of Rochester, Rochester, NY 14627}
\affiliation{Center for Coherence and Quantum Optics, University of Rochester, Rochester, NY 14627}
\email{swadood@ur.rochester.edu}
\author{Eric D. Black}
\affiliation{Department of Physics, California Institute of Technology, MC 264-33, Pasadena, CA 91125}
\email{blacke@caltech.edu}
\author{Peter W. Milonni}
\affiliation{Theoretical Division, Los Alamos National Laboratory, Los Alamos, New Mexico 87545} 
\affiliation{Department of Physics and Astronomy, University of Rochester, Rochester, NY 14627}
\email{peter\_milonni@comcast.net} 

\begin{abstract}
A proposed experiment to test whether space is discretized [J. D. Bekenstein, Phys. Rev. D {\bf 86}, 124040 (2012); Found. Phys. {\bf 44}, 452 (2014)] is based on the supposed impossibility of an incident photon causing a displacement of a transparent block by less than the Planck length. An analysis of the quantum and thermal jitter of the block shows that it greatly diminishes the possibility that the experiment could reveal Planck-scale signals.   
\end{abstract}
\keywords{Planck length, zero-point energy, zero-point motion, harmonic oscillator, quantum jitter}
\maketitle
\section{introduction}
Bekenstein \cite{bek,bek2} has made the remarkable suggestion that it might be possible to test Wheeler's notion of ``quantum foam" \cite{wheeler} in a tabletop experiment ideally involving a single photon incident on a suspended, transparent block. If the transmission of the photon would displace the block by less than the Planck length $L_p$, and if displacements less than $L_p$ cannot occur, then conservation of momentum prevents transmission of the photon. The photon must therefore be reflected or absorbed. If its frequency is far removed from any absorption frequency of the block, the photon must evidently be reflected, with the change in its momentum taken up by the block. Reflection probabilities greater than expected from the Fresnel reflection coefficient would therefore serve as evidence that displacements smaller than $L_p$ cannot occur.

In the proposed experiment \cite{bek,bek2}, a photon of frequency $\om$ is incident on a block of mass $M$ and real refractive index $n$. The block is assumed to be suspended in a vacuum by a fiber of length $\ell$. The photon momenta inside and outside the block are $\hbar\om/nc$ and $\hbar\om/c$, respectively, so that if the photon traverses the block length $L$ before exiting there is a transfer of momentum $P=(1-1/n)\hbar\om/c$ to the block. The transit time for the photon to cross the block is $\tau=nL/c$, so the net displacement of the center of mass is \cite{nandor} 
\be
\delta x=(P/M)nL/c=(\hbar\om/Mc^2)(n-1)L .
\ee
 If this is impossible for $\delta x<L_p=\sqrt{\hbar G/c^3}\approx 1.6\times 10^{-35}$ m, the photon cannot be transmitted and must be reflected. Using Bekenstein's experimental parameters ($M=1.5\times 10^{-4}$ kg, $ L=10^{-3}$ m,\ $ \ell = 0.1$ m, $ n=1.6$, $ \hbar\om=2.8$ eV), we find $\delta x=1.2L_p$ suggesting that parameters can be chosen that might make this experiment possible in principle.
 
 McDonald has argued that no Planck-scale physics can be revealed by measuring only the transmission (or reflection) coefficient in the proposed experiment, since the experiment does not measure $\delta x$, which therefore cannot be said to have a displacement smaller than $L_p$ \cite{kirk}. Here we raise a very different concern regarding the feasibility of the proposed experiment---the inescapable jitter of the block.
 
Bekenstein assumes, as we will, that the glass block can be treated as a rigid body. Thermal jitter of the block due to impacts from surrounding photons or molecules can be reduced by carrying out the experiment at sufficiently low temperatures $T$ and pressures $P$, and Bekenstein has carefully shown that it is possible thereby to effectively eliminate this source of center-of-mass fluctuations \cite{bek}.  

Bekenstein analyzed the experiment in the rest frame of the block, so that any motion of the glass due to its suspension by a fiber could essentially be ignored. As he states, ``the motion of the  block [center of mass (c.m.)] that we speak of does not include the effect of the force from the fiber. Of course, this last does play a part in establishing the motion of the block at any moment; such motion, however, is not considered here because we work in the block c.m.'s Lorentz frame at the moment of photon ingress \cite{bek}." Because of the low fundamental frequency of this oscillator, about 10 Hz, and the short transit time, $\tau \approx 5$ ps, this assumption seems reasonable.  However, when we analyze the suspended glass as a quantum harmonic oscillator, it becomes apparent that the jitter, relative to the Planck length, is so great that no Lorentz frame exists for the required amount of time.

In the following section we analyze the system as a quantum harmonic oscillator \cite{tsip} with arbitrary $\hbar \omega/k_BT$ and calculate the position and momentum fluctuations associated with internal damping of the mechanical system, an effect not considered by Bekenstein. We consider temperatures $T=0$ K (zero-point motion) as well as  $T>\hbar\omega/k_B$. Although thermal effects for $T=1$ K result in an rms jitter some 5 orders of magnitude greater than that for zero-point jitter, the nonreducible zero-point jitter alone makes success of the experiment as proposed very unlikely. We regard the computed jitter as a real-time phenomenon \cite{wine}, and not solely a reflection of the measured variability in a set of identically prepared systems. In Section \ref{sec:sec3} we consider the effect of the momentum transfer on the ground state of the oscillator and show that its effect on the ground-state jitter is  completely negligible. In Section \ref{sec:sec4} we take a different perspective on the proposed experiment; we estimate the probability that the momentum of the block encountered by the incident photon is sufficiently small that the displacement of the block during the photon transit would be about a Planck length, in which case the proposed experiment to test for a Planck-scale signal might be feasible. Our conclusions are briefly summarized in Section \ref{sec:sec5}.

\section{Langevin Noise Analysis of Jitter}
The block is suspended by a fiber of length $\ell$ and acts as a harmonic oscillator with a natural oscillation angular frequency  $\om_{0}=\sqrt{g/\ell}$ for small angular displacements from the vertical. For Beckenstein's parameters, $\omega_{0} \approx 9.9$ rad/s and the corresponding frequency is about $1.6$ Hz.  

Consider the fluctuations in the displacement of the block associated with internal damping in the mechanical system, which Saulson has analyzed classically using a model based on a complex spring constant \cite{5,6}. The damping characterized by the complex spring constant implies, by the fluctuation-dissipation theorem, that there must be fluctuations in the center of mass ($x$) .  For our purposes it is convenient to work with an equation of motion given approximately, as described by Saulson, by
\begin{equation}
    \ddot x +\omega_{0}^2x +i\omega_{0}^2\phi x=\frac{1}{M}F(t),
\end{equation}
where $i\omega_{0}^2\phi$ is the complex spring constant \cite{5} and $F(t)$ is a Langevin noise force. $\phi$ may be assumed to be very small, say $\phi\approx 10^{-6}$ for a weakly damped system \cite{size}. The noise is assumed to result from the coupling of the system to a thermal reservoir of oscillators with raising and lowering operators $a$ and $a^{\dagger}$:
\begin{equation}
    F(t)=C\left[\int_{0}^{\infty}d\omega a(\omega)e^{-i\omega t} + \int_{0}^{\infty} a^{\dagger}(\omega) e^{i \omega t}\right],
\end{equation}
with \begin{equation}
    [a(\omega),a(\omega^{\prime})]=0,\ [a(\omega),a^{\dagger}(\omega^{\prime})]=\delta(\omega-\omega^{\prime}) .
\end{equation}
The constant $C$ is defined as 
\begin{equation}
    C=\Big(\frac{M\hbar \omega_{0}^{2} \phi}{\pi}\Big)^{1/2} .
    \end{equation}
This will be seen to be consistent with expressions obtained by Saulson \cite{5}, and is the form implied by the fluctuation-dissipation theorem.

The steady-state solutions for $x$ and $p=M\dot{x}$ are 
\be
    x(t)=\frac{C}{M} \left[ \int_{0}^{\infty} \frac {d\omega a(\omega) e^{-i\omega t}}{\omega_{0}^2-\omega^{2} + i \omega_{0}^2 \phi}+\int_{0}^{\infty} \frac{d\omega a^{\dagger}(\omega) e^{i \omega t}}{\omega_{0}^2-\omega^{2} + i \omega_{0}^2 \phi} \right]
\ee
and
\bea
p(t)&=&iC\Big\{-\int_0^{\infty}\frac{d\om \om a(\om)e^{-i\om t}}{\om_0^2-\om^2+i\om_0^2\phi}\nonumber\\
&&\mbox{}+
\int_0^{\infty}\frac{d\om \om a^{\dagger}(\om)e^{i\om t}}{\om_0^2-\om^2+i\om_0^2\phi}\Big\}.
\label{pos}
\eea
For the reservoir maintaining thermal equilibrium at temperature $T$, 
\be
\la a^{\dagger}(\om)a(\om^{\prime})\ra=\delta(\om-\om^{\prime})\frac{1}{e^{\hbar\om/k_BT}-1},
\label{reft1}
\ee
\be
\la a(\om)a^{\dagger}(\om^{\prime})\ra=\delta(\om-\om^{\prime})\Big[1+\frac{1}{e^{\hbar\om/k_BT}-1}\Big],
\label{reft2}
\ee
and
\be
\la a(\om)a(\om^{\prime})\ra=0,
\ee
from which we obtain, for example,
\bea
\la x^2(t)\ra&=&\frac{C^2}{M^2}\int_0^{\infty}d\om\Big\{1+
\frac{1}{e^{\hbar\om/k_BT}-1}+\frac{1}{e^{\hbar\om/k_BT}-1}\Big\} \nonumber\\
&&\mbox{}\times\frac{1}{(\om^2-\om_0^2)^2+\om_0^4\phi^2},
\eea
and
\bea
\la p(t_1)p(t_2)\ra&=&C^2\int_0^{\infty}d\om\Big\{\Big[1+\frac{1}{e^{\hbar\om/k_BT}-1}\Big]e^{i\om(t_2-t_1)}\nonumber\\
&+&\frac{e^{-i\om(t_2-t_1)}}{e^{\hbar\om/k_BT}-1}\Big\}
 \frac{\om^2}{(\om^2-\om_0^2)^2+\om_0^4\phi^2}.
\eea
For the 100 mm fiber length assumed by Bekenstein, $\om_0=\sqrt{g/\ell}\approx 10$ rad/s and $k_BT\gg\hbar\om_0$  for realizable temperatures. Then $k_BT\gg\hbar\om$ for frequencies that contribute significantly to the integrals, and \cite{note}
\be
\la x^2(t)\ra\cong \frac{2\om_0^2\phi k_BT}{\pi M}\int_0^{\infty}\frac{d\om}{\om[(\om^2-\om_0^2)^2+\om_0^4\phi^2]},
\label{eq100}
\ee
which is equivalent to Eq. (16) of Saulson's paper \cite{5}, and
\be
\la p(t_1)p(t_2)\ra\cong \frac{2\om_0^2\phi Mk_BT}{\pi}\int_0^{\infty}\frac{d\om\om\cos\om(t_1-t_2)}{(\om^2-\om_0^2)^2+\om_0^4\phi^2}.
\label{pcorr}
\ee
 
We define the variance
\be
\la \delta x^2(\tau)\ra=\frac{1}{M^2}\int_0^{\tau}dt_1\int_0^{\tau}dt_2\la p(t_1)p(t_2)\ra
\label{vardef}
\ee
in the position of the block over the interval of the photon transit time $\tau\approx 5$ ps and obtain, from (\ref{pcorr}),
\bea
\la \delta x^2(\tau)\ra&=&\frac{2\om_0^2\phi k_BT}{\pi M}\int_0^{\infty}\frac{d\om\om}{(\om^2-\om_0^2)^2+\om_0^4\phi^2}\nonumber\\
&&\mbox{}\times\int_0^{\tau}dt_1\int_0^{\tau}dt_2\cos\om(t_1-t_2)\nonumber\\
&=&\frac{2\om_0^2\phi k_BT}{\pi M}\tau^2\int_0^{\infty}\frac{d\om\om}{(\om^2-\om_0^2)^2+\om_0^4\phi^2}\nonumber\\
&&\mbox{}\times\frac{\sin^2\frac{1}{2}\om\tau}{(\frac{1}{2}\om\tau)^2}.
\eea
Since $\phi\ll 1$, and $\tau$ is extremely small compared to $\omega_{0}^{-1}$, the sinc function is very broad compared to the function in the integrand that peaks at $\om=\om_0$, and so
\bea
\la \delta x^2(\tau)\ra&\cong&\frac{2\om_0^2\phi k_BT}{\pi M}\tau^2\frac{\sin^2{\frac{1}{2}\om_0\tau}}{(\frac{1}{2}\om_0\tau)^2}\nonumber\\
&&\mbox{}\times\int_0^{\infty}\frac{\om d\om}{(\om^2-\om_0^2)^2+\om_0^4\phi^2}\nonumber\\
&\cong&\frac{k_BT}{M}\tau^2,
\label{varint}
\eea
where we have approximated the sinc function multiplying the integral by 1, since $\om_0\tau \approx 10^{-11}$. This result for $\la \delta x^2(\tau)\ra$ is just $(\om_0\tau)^2$ times the thermal equilibrium value $k_BT/M\om_0^2$ of $\la x^2\ra$ given by the equipartition theorem. The parameter $\phi$ characterizing the internal damping does not appear because the time $\tau$ is too short for any damping to occur. 

The rms displacement $\la \delta x^2(\tau)\ra^{1/2}$ over the transit time $\tau$ due to internal damping and the consequent fluctuations is the thermal velocity $(k_BT/M)^{1/2}$ times $\tau$,  and is much greater than the Planck length. For the parameters assumed
by Bekenstein, for example, $[\la\delta x^2(\tau)\ra]^{1/2}\approx 1.6\times 10^{-21}\sqrt{T(K)}$ m $\approx  10^{14}\sqrt{T(K)}L_p$. The thermal jitter of the
block thus makes it impossible, as a practical matter, ``to sidestep the onerous requirement of localization of a probe on the Planck length scale" \cite{bek} for any realizable temperature.

A fundamental limit on the localization of the probe is the center-of-mass fluctuation that persists even as $T\rightarrow 0$. The variance $\la \delta x^2(\tau)\ra_{\rm zp}$ in this zero-point jitter can be derived as above in the limit in which $\phi$ and $T$ approach 0, or more simply from the solution of the Heisenberg equation of motion for the momentum $p(t)$:  $p(t)=p(0)\cos\om_0t-M\om_0x(0)\sin\om_0t$. Then, since $\la x^2(0)\ra=\hbar/2M\om_0$,
$\la p^2(0)\ra=M\hbar\om_0$, and $\la x(0)p(0)\ra=-\la p(0)x(0)\ra=i\hbar/2$ for the ground state of the oscillator, it follows from (\ref{vardef}) that
\be
\la\delta x^2(\tau)\ra_{\rm zp}=\frac{\hbar\om_0}{2M}\frac{\sin^2\frac{1}{2}\om_0\tau}{(\frac{1}{2}\om_0)^2}\cong \frac{\hbar\om_0}{2M}\tau^2
\ee
for $\om_0\tau\ll 1$, which could have been deduced from (\ref{varint}) by simply replacing the thermal energy $k_BT$ by the zero-point energy $\frac{1}{2}\hbar\om_0$. For the parameters assumed by Bekenstein, $[\delta x^2(\tau)_{\rm zp}]^{1/2}\approx 6\times 10^7L_p$.  The zero-point variance adds to that estimated from (\ref{varint}) and is much smaller, but would by itself make it very difficult to probe the Planck scale in the proposed experiment.

Because of the low frequency of the oscillator, $9.9 $ rad/s, one might argue that, despite this large variance $\la\delta x^2(\tau)\ra$, it might be possible that the jitter has, for a short time at least, a constant velocity component in some inertial frame.  To investigate this possibility we can simply compute the variance in velocity $\la\delta v^2(\tau)\ra$ following the same procedure as before. We define

\begin{eqnarray}
&&\la \delta \text{v}^{2}(\tau) \ra =\frac{1}{M^{2}}\int_{0}^{\tau
}dt_{1}\int_{0}^{\tau }dt_{2}\big\la\dot{p}(t_1)\dot{p}(t_2)\big\ra\nonumber\\
&=&\int_{0}^{\tau }dt_{1}\int_{0}^{\tau }dt_{2}\int_{0}^{\infty }d\omega 
\frac{2\omega _{0}^{2}\phi k_BT}{\pi M}\frac{\omega ^{3}\cos \omega
(t_{1}-t_{2})}{(\omega ^{2}-\omega _{0}^{2})^{2}+\omega _{0}^{4}\phi ^{2}}.\nonumber\\
\end{eqnarray}
After doing the integrations, making the same approximations as before, and retaining the dominant terms, we find
\begin{equation}
\la\delta\text{v}^{2}(\tau )\ra=\frac{k_BT}{M}(\omega _{0}\tau )^{2}.
\end{equation}
The rms variation in velocity $[\la\delta\text{v}^{2}(\tau )\ra]^{1/2}$  is the rms thermal
velocity $(k_BT/M)^{1/2}$ times $(\omega _{0}\tau )$ .
In terms of $\la\delta $x$^{2}(\tau )\ra$,
\begin{equation}
\la\delta \text{v}^{2}(\tau )\ra=\la\delta \text{x}^{2}(\tau )\ra\omega _{0}^{2}.
\end{equation}
For a temperature of 1 K, $\left( \la\delta \text{v}^{2}(\tau )\ra\right) ^{1/2}$ = $9.9\times 10^{14} L_{P}/s$ = 
$5250  L_P/\tau$, which suggests that we cannot simply assume an inertial frame. 

Equation (21) holds also for the zero-point fluctuations. For our parameters,  $\left( \la\delta \text{v}^{2}(\tau )\ra_{zp}\right) ^{1/2} $ is approximately $0.03$ $L_P/\tau$, where we have chosen units most appropriate to the experiment. This zero-point variance in the velocity is small enough to suggest that there is a Lorentz frame  in which the block is at rest during the transit time. But this would require a temperature of approximately $10^{-11}$ K.  For currently feasible experimental conditions, it is apparent, statistically at least,  that we cannot assume that the motion during the transit time occurs in an inertial frame.

It is important in this connection to distinguish between the thermal fluctuations we have considered and those analyzed by Bekenstein \cite{bek,bek2}. 
The latter are ``maintained by collisions of ambient gas atoms or molecules and thermal photons with the block" \cite{bek}, and are found by 
Bekenstein to be ``the most troublesome source of noise" \cite{bek};  at any realistic temperature this jitter would seem to completely swamp the Planck length and preclude a successful experiment. However, collisions between the block and the surrounding gas atoms could be so rare, at sufficiently low pressures, that
the block is free of any jitter between atom hits. Bekenstein makes a simple estimate of the
probability $\Pi$ of a hit  during the photon transit time and finds
\be
\Pi=nL^2L_1P\sqrt{\frac{3}{mc^2k_BT}},
\ee
where P is the pressure, $L^2$ is the surface area of the block, $L_1$ is the block length traversed by the
photon, and $m$ is the mass of a helium atom (helium assumed to be the ambient gas). The experiment could be
feasible if $\Pi\ll 1$, say $\Pi\leq .01$. For $T=1$ K, $n=1.6$, $L_1=1$ mm, and $P=10^{-11}$ Pa , this requires
\be
L\leq 3 \ {\rm cm}.
\ee
 The experiment as envisaged by Bekenstein could therefore work if the pressure is sufficiently low and the block is sufficiently small
that in some time intervals the block is free of any thermal jitter due to molecular collisions. In those intervals the block would move with uniform velocity and be at rest in some Lorentz frame. 

The fluctuations we have considered, however, are connected by the fluctuation-dissipation theorem to {\sl internal} damping, specifically the dissipation of elastic energy of the fiber supporting the pendulum against gravity \cite{size}. For these thermal fluctuations at the temperature $T$ of the support system and the surroundings there are no time intervals in which the block can be supposed to have a constant velocity during the photon transit time.  The jitter due to these
thermal fluctuations cannot be avoided by simply pumping out enough gas. 

\section{Effect of the Momentum Impulse on the Ground State of the Oscillator}\label{sec:sec3}  

The zero-point jitter (18) of the unperturbed block would alone make this experiment very difficult. In order to determine how the ground-state jitter might be altered by the momentum transfer from the photon, we consider the Hamiltonian
\begin{equation}
    H=\hbar \omega_{0} (a^{\dagger} a+\frac{1}{2}) + \frac{1}{M}f(t)p ,
\end{equation}
where $f(t)=P\big[\Theta(t) - \Theta(t-\tau)\big]$ and $\Theta(t)$ is the Heaviside step function. $P$ represents the momentum transfer while the photon is in the block during the transit time $\tau$. 
We will evaluate $\la\delta x^2(\tau)\ra_{\rm Pzp}$ using the Heisenberg equation of motion $i\hbar\dot a=[a,H]$ that follows from this Hamiltonian. The subscript Pzp refers to the ground state when the momentum impulse $P$ is present. For the momentum and position operators during this time we obtain
\begin{equation}\label{eqn::momentumheisenberg}
    p(t) = i \sqrt{\frac{M\hbar \omega_{0}}{2}}(a^{\dagger}(0)e^{i \omega_{0} t}-a(0)e^{-i \omega_{0} t})+P( \cos{\omega_{0} t}-1)
\end{equation}
and
\begin{equation}\label{eqn::positionheisenberg}
  x(t) = \sqrt{\frac{\hbar}{2 M \omega_{0}}}(a^{\dagger}(0)e^{i \omega_{0} t}+a(0)e^{-i \omega_{0} t})+\frac{P}{M \omega_{0}}\sin{\omega_{0} t} .  
\end{equation}

The first terms in these expressions are of course just the position and momentum operators, respectively, for the unperturbed oscillator. The remaining terms are due to the momentum impulse, and are the only terms that survive when we take expectation values for the ground state: $\langle p(\tau) \rangle_{Pzp} =P (\cos{\omega_{0}\tau}-1)\approx P \omega_{0}^2\tau^{2}/2<<P $.  Similarly $\langle x(\tau) \rangle_{Pzp} =\frac{P}{M \omega_{0}} (\sin{\omega_{0}\tau})\approx P\tau/M\approx L_{P}$ by experimental design. For the variance $\la\delta x^2(\tau)\ra_{\rm Pzp}$, we use (15) and (21) to obtain
\begin{equation}
    \la\delta x^2(\tau)\ra_{\rm Pzp}=\la\delta x^2(\tau)\ra_{\rm zp} + (\frac{P}{M \omega_{0}})^{2}(\sin{\omega_{0} t} -\omega_{0} \tau)^2 .
\end{equation}
The first term is the result (18) when no impulse was present, and the second term is the correction for $P\neq 0$. For the assumed experimental parameters, $\omega_{0} \tau \approx  10^{-11}$ and the correction is 
\begin{equation}
     \la\delta x^2(\tau)\ra_{\rm Pzp} - \la\delta x^2(\tau)\ra_{\rm zp} \approx \frac{P^2 \omega_{0}^4 t^6}{36 M^2}\approx 10^{-113} \ {\rm m}^2 . 
\end{equation}
Thus, as implicitly assumed above, the effect of the impulse $P$ on the zero-point variance $\la\delta x^2(\tau)\ra_{\rm zp}$ is negligible.
\\

\section{Probability of experimental success based on momentum restrictions}\label{sec:sec4}

As Bekenstein noted, the experiment ``cannot employ a macroscopic light pulse instead of a single photon. In the former case the pulse is always partially reflected
back, and the resulting recoil of the block imparts to it a constant velocity leading ultimately to unlimited translation" \cite{bek}. We have argued that the unavoidable fluctuations in the position of the block will, on average, result in its translation by much more than $L_p$ during the transit of a single photon; then it is unlikely that the photon will experience any anomalous reflection that would be expected without these fluctuations. It is natural to ask what is the likelihood of observing anomalous reflection in repeated single-photon measurements. For a rough estimate, let us assume that the largest block momentum 
$p_{\rm max}$ that results in a displacement less than $L_p$ during a photon transit is given by $(p_{\rm max}/M)\tau=L_p$. Then we identify a probability of success for observing an anomalous reflection in each of a series of repeated single-photon observations, based on the ground-state momentum wave function 
$\phi_0(p)$ for the block, as 
\bea\label{eqn::wigner groundstate}
\mathcal{P}_{\rm zp}&=&\int_{-p_{\rm max}}^{p_{\rm max}}dp|\phi_0(p)|^2\nonumber\\
&=&\big(\frac{1}{\pi M\hbar\om_0}\big)^{1/2}\int_{-p_{\rm max}}^{p_{\rm max}}dpe^{-p^2/M\hbar\om_0}\nonumber\\
&\approx&\big(\frac{4M}{\pi\hbar\om_0}\big)^{1/2}\frac{L_p}{\tau}\approx 1.3\times 10^{-9},
\label{est1}
\eea
\\
since $p_{\rm max}\ll (M\hbar\om_0)^{1/2}$. This is the probability that the block momentum is less than $|p_{max}|$ and defines an upper limit, set by zero-point jitter, to the probability of an observation of anomalous reflection.  For the much larger, thermally-driven center-of-mass fluctuations, we estimate, similarly, 
\begin{align}\label{eqn::wigner thermal}
\mathcal{P}_T\approx (\frac{\ 2M}{\pi k_BT})^{1/2}\frac{L_p}{\tau} \approx \frac{8\times 10^{-15}}{\sqrt{T(K)}}.
\end{align}
\\which gives a probability of success about 5 orders of magnitude smaller than (\ref{est1}) for $T=1$ K.
\par
This result can be also understood from a purely classical perspective. The change in position $\delta x_{\tau}$ of an oscillator relative to $L_{p}$ after the transit time $\tau$ can be approximated as 
\begin{align}
   \frac{\delta x_{\tau}}{L_{p}}= \frac{\tau p_{0}}{ML_{p}}+\frac{P\tau}{ML_{p}}.
\end{align}
If the photon strikes the oscillator when it is \textit{exactly} at a turning point, i.e., $p_{0}=0$, we would get a displacement from only the second term, which is of the order of $L_{p}$ by design. However, if the oscillator is $NL_{p}$ away from the classical turning point $x_{t}$ at the time of impact, then the block momentum is not zero and 
\begin{align}\label{eqn::classicaloscillator}
    \frac{\delta x_{\tau}}{L_{p}}=\frac{\tau\omega\sqrt{2x_{t}NL_{p}}}{L_{p}}+\frac{P\tau}{ML_{p}}.
\end{align}
The first term is the excess displacement due to the oscillator's initial energy. Based on Bekenstein's parameters, the first term on the right side of Eq. (\ref{eqn::classicaloscillator}) is approximately 0.3,10 and 300 for $N=1,10^{3}$, and $10^{6}$ respectively. This shows that for a successful experiment, the photon needs to strike the oscillator when it is near $x_{t}$ to the accuracy of $L_{p}$, or equivalently, when it's momentum is near zero, accurate to order of $p_{max}$. The corresponding quantum mechanical probabilities for such a ``timed hit" are expressed by Eqs. (\ref{eqn::wigner groundstate}) and (\ref{eqn::wigner thermal}).

In a less simplistic classical model the photon could be described as a wavepacket with a physical extent of perhaps several wavelengths. The wavelength is $0.4 $ $\mu$m so its extent might be  2-3 $\mu$m. This is much smaller than the thickness assumed for the block (1000 $\mu$m), so that the photon transit time in this model is well defined \cite{nandor}, but its spatial extent is about $10^{28}$ $L_{p}$, which makes it very difficult to imagine that the photon arrival occurs at the instant when the glass momentum $p_0$ is zero.  
\vspace{0.3cm}
\section{Conclusions}\label{sec:sec5}
In analyzing his ingenious proposal, Bekenstein concluded that the problem of thermal fluctuations of the block's position due to collisions with an ambient gas could be circumvented by conducting the experiment at sufficiently low temperatures and pressures. 
We have argued that the zero-temperature jitter of the block, which was not addressed by Bekenstein, appears to make it impossible, for realistic experimental parameters, ``to sidestep the onerous requirement of localization of a probe on the Planck length scale" \cite{bek}. Success of the proposed experiment in revealing Planck-scale signals appears to be even more unlikely when we take into account the fluctuations that must, according to the fluctuation-dissipation theorem, accompany the internal damping of the pendulum system. 
  
\section*{acknowledgement} We thank Lowell S. Brown, Richard J. Cook, Ashok Das, Joseph H. Eberly, Adrian C. Melissinos, Emil Mottola, and Peter R. Saulson for their helpful comments. We especially thank Rainer Weiss for so generously sharing his insights relating to the Bekenstein proposal and for pointing out to us the importance of thermal noise due to damping processes even at very low temperatures and pressures, which we had previously neglected by considering only zero-point fluctuations. Research by S. A. Wadood was supported by the Army Research Office under grant no. W911NF1610162.


\begin{thebibliography}{99}
\bibitem{note} For a discussion of the formal divergence of (13), see Reference \cite{5}. Since the integrand peaks around $\omega = \omega_{0}$, the integral is finite for, and insensitive to, any sufficiently small lower cutoff, as is easily verified numerically.
\end{thebibliography}
\end{document}